\documentclass[amsmath,preprintnumbers,10pt,twocolumn,prl]{revtex4-1}
\usepackage{amsmath,amsfonts,bm}
\usepackage{graphicx}
\usepackage{color}
\usepackage{subfigure}
\usepackage{verbatim}

\begin{document}
\title{Putting water on a lattice: The importance of long wavelength density fluctuations in theories of hydrophobic and interfacial phenomena.}
\author{Suriyanarayanan Vaikuntanathan$^{1}$ and Phillip L Geissler$^{1,2}$}
\affiliation{$^1$Material Sciences Division, Lawrence Berkeley National Lab, Berkeley, CA 94720. \\ 
$^2$Department of Chemistry, University of California, Berkeley, CA 94720.}
\begin{abstract}
The physics of air-water interfaces plays a central role in modern
theories of the hydrophobic effect. Implementing these theories,
however, has been hampered by the difficulty of addressing
fluctuations in the shape of such soft interfaces.  We show that this
challenge is a fundamental consequence of mapping long wavelength
density variations onto discrete degrees of freedom.  Drawing from
studies of surface roughness in lattice models, we account for the
resulting nonlinearities simply but accurately.  Simulations show that
this approach captures complex solvation behaviors quantitatively.

\end{abstract}
\maketitle 

The fluctuating roughness of liquid-vapor interfaces spans a wide
range of length scales: from the mesoscopic, where 
the coarse view of 
capillary wave
theory 
is appropriate~\cite{Weeks1977}, to the
microscopic~\cite{Chandler1993,Hummer1996}, where molecular
considerations are essential.  Computer simulations have demonstrated
that these topographical fluctuations can impact an equally broad
spectrum of physical responses, with important implications for
behaviors of modern interest in biophysics, chemical physics, and
materials science including:  
binding of ligands to hydrophobic protein cavities~\cite{Setny2013},
self assembly of nanoparticles at interfaces~\cite{Cavallaro2011}
and the affinity of diverse solutes for the liquid's boundary
~\cite{Vaikuntanathan2013,Otten2012,Levin2009}.
Lum, Chandler, and Weeks 
(LCW)
developed a comprehensive conceptual
framework for linking such solvation phenomena to fluctuations in the
liquid's microscopic density field~\cite{Lum1999}, to
which surface roughness clearly contributes.
But the
corresponding theory has been thoroughly explored only within
mean-field approximations for long-wavelength
response~\cite{Lum1999}. Attempts to
simultaneously address fluctuations at fine and coarse scales have
been hampered by difficulties associated with faithfully representing
long-wavelength modes in a statistical mechanics
model~\cite{ReintenWolde2001,Varilly2011,Patel2012} and as a result have suffered from unphysical degeneracies~\cite{ReintenWolde2001,Varilly2011} or else from the need to introduce
numerous parameters that are poorly constrained by available
data~\cite{Varilly2011,Patel2012}.
\begin{figure}[tbp]
                      \includegraphics[width=0.42\textwidth]{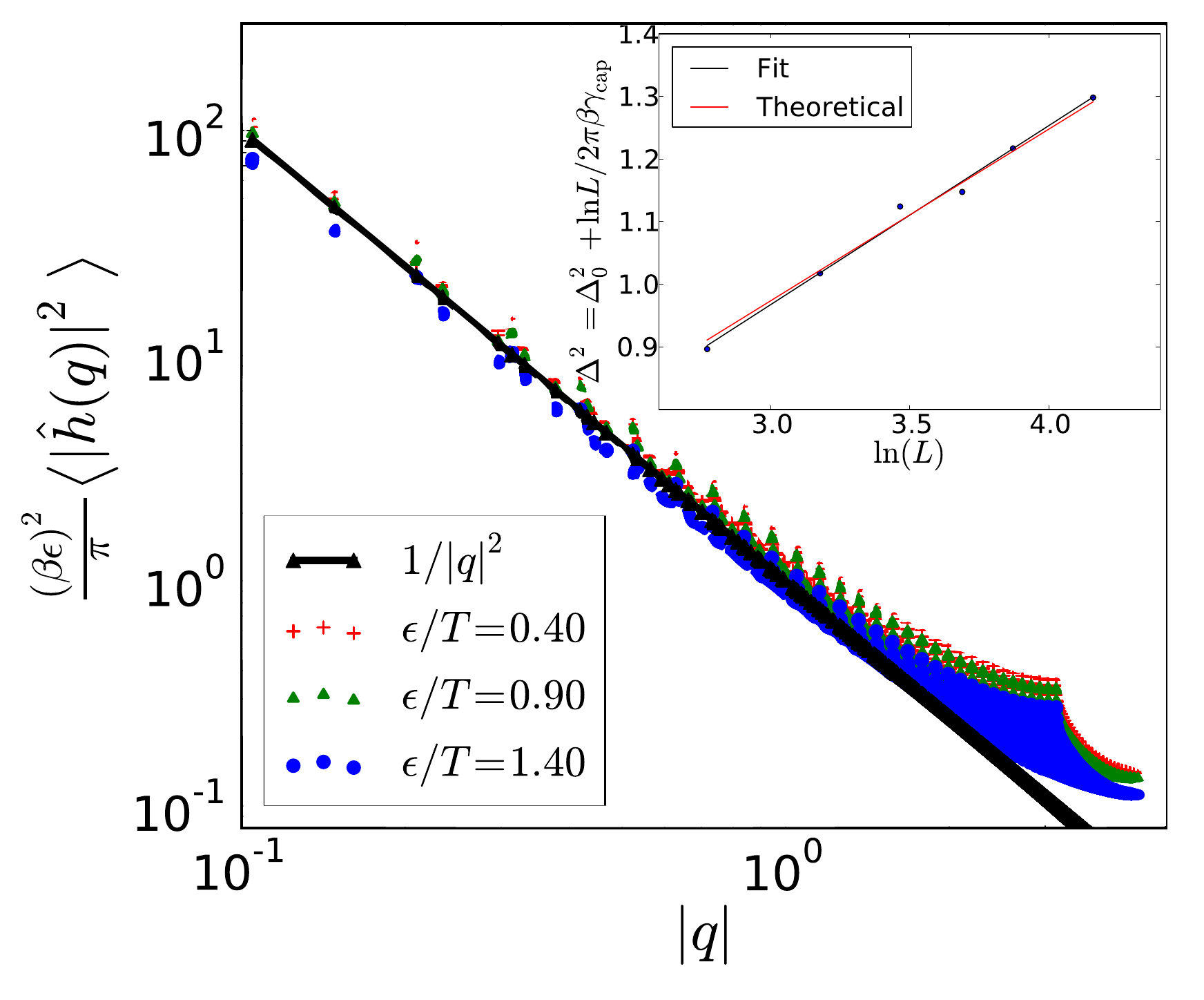}
            \caption{
(Log-Log) Plots of $(\beta\epsilon)^{2}\langle |\hat h (q)|^2\rangle/\pi$, where $\hat h(q)$ 
denotes
the Fourier modes of interfacial fluctuations in a Solid-on-Solid 
model.  
Over a fairly large range of 
the lattice coupling strength
$\epsilon/T$, this plot demonstrates that $(\beta\epsilon)^{2}\langle |\hat h (q)|^2\rangle/\pi\approx 1/|q|^2$ for small values of $|q|$, thus verifying the scaling predicted in Eq.~\ref{eq:centralresult1}. 
Inset: 
Capillary fluctuations of the lattice gas, at $\epsilon/T=1.35$.
The squared interfacial thickness $\Delta^2$ determined from
simulations, which reflects on the extent of surface roughness,
depends logarithmically on the lateral dimension $L$,
as predicted by capillary wave theory.
The proportionality coefficient for this dependence indicates the
surface tension $\gamma_{\rm cap}$. The value extracted by fitting
simulation results to the expected form
agrees well with predictions of 
Eq.~\ref{eq:centralresult1}. Details of calculations in SM~\cite{Suppref1}.}
 \label{fig1:SOSlattice}
\end{figure}
In this Letter we show that previous difficulties in modeling 
coarse variations in liquid density~\cite{ReintenWolde2001,Varilly2011}
reflect the rich statistical physics of discretely fluctuating
surfaces, whose relevance in this context has not been elaborated. In
particular, the component of the density field that
varies slowly in space is naturally described in numerical approaches
by a lattice model. This choice of a discrete representation
introduces profound nonlinearities, which can
cause decoupling of notionally equivalent measures of surface
tension. In extreme cases these nonlinearities can even drive a phase
transition from a rough to a quiescent state~\cite{Weeks1973}, which lacks
the long-wavelength fluctuations 
altogether.

\begin{figure*}[tbp]
     \subfigure[]{
             
                \includegraphics[scale=0.32]{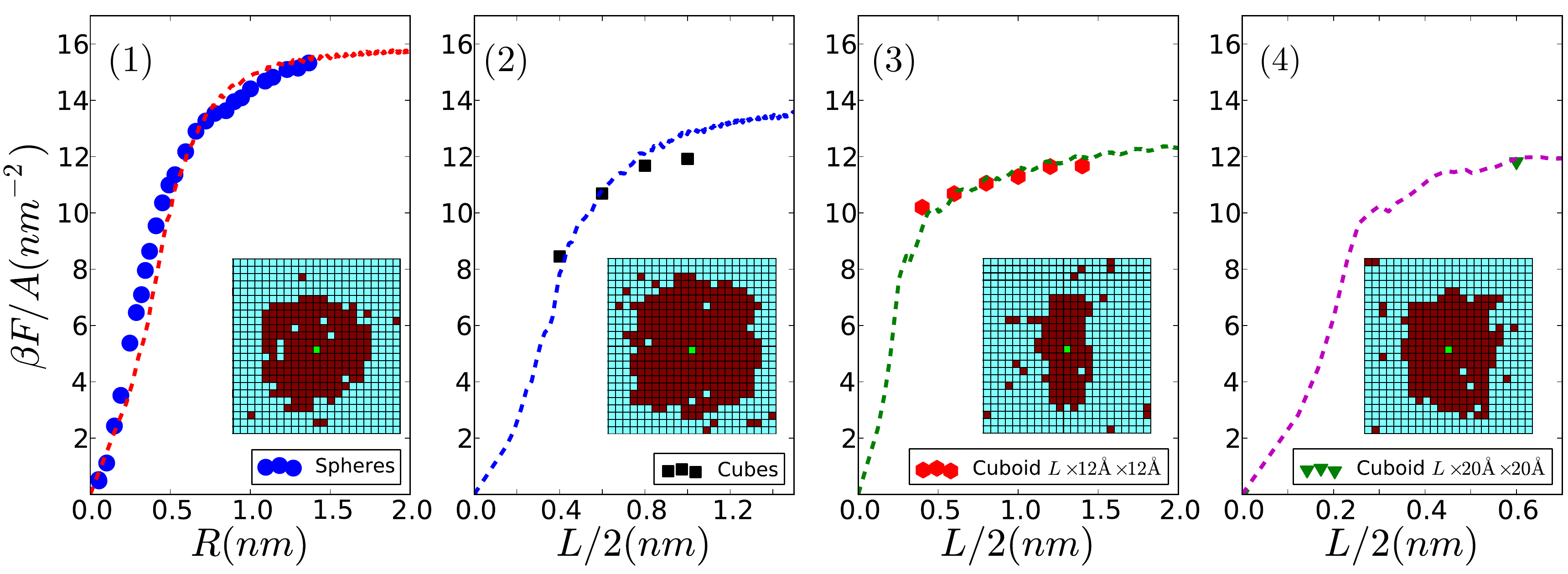}
                 \label{fig2:results}}
     \subfigure[]{
                \includegraphics[scale=0.28]{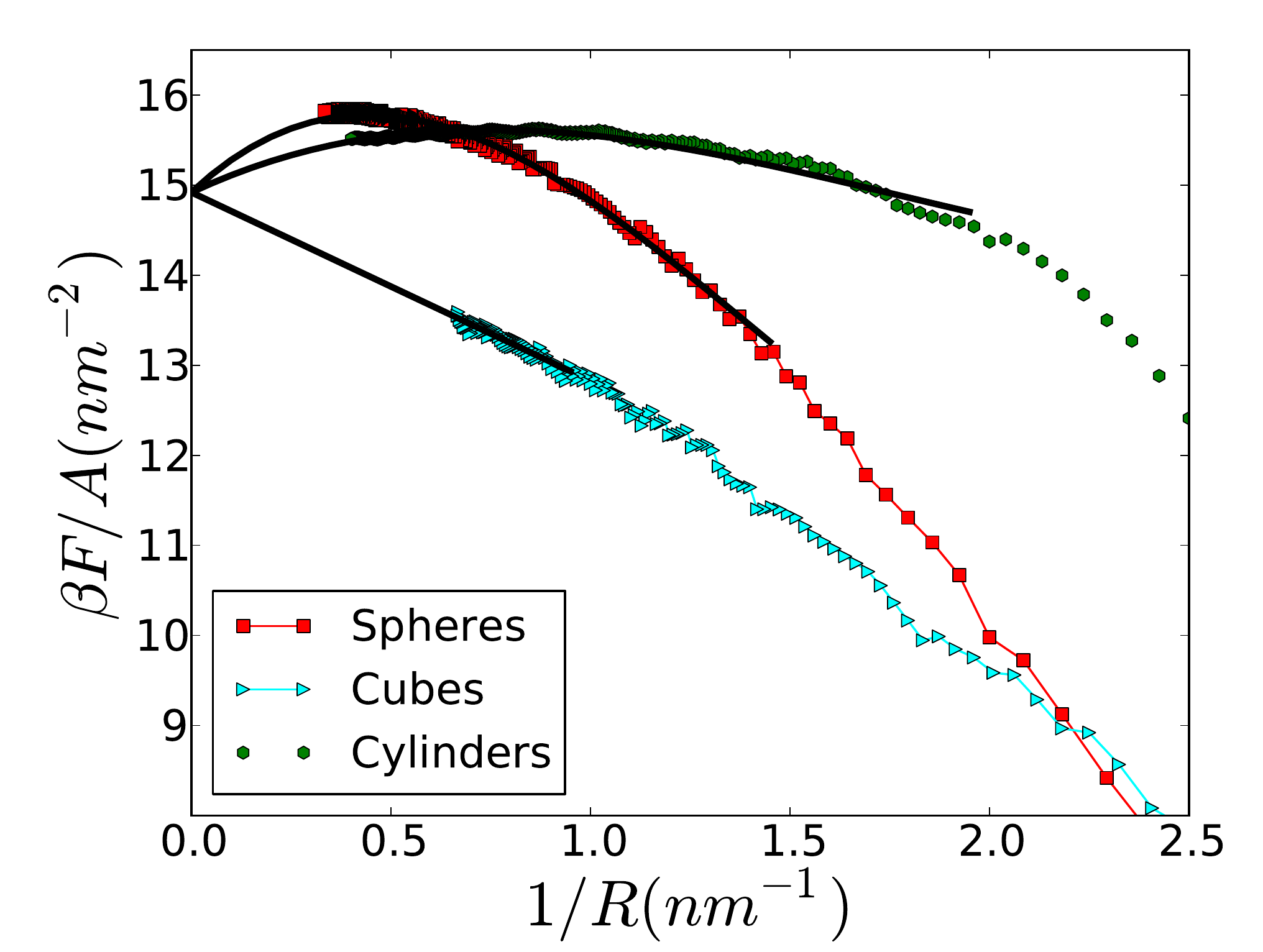}
               \label{fig3:results}}
                \caption{(a) Estimates of solvation free energies per unit area of ideal hydrophobic spheres, cubes, and cuboids of various sizes obtained from both the coarse grained lattice model (lines), Eq.~\ref{eq:LCW}, and atomistic simulations of SPC/E water (points). (Inset) Cross sections of snapshots of lattice gas simulations using Eq.~\ref{eq:LCW}. The slowly varying density field $n_i$ is predominantly zero in regions occupied by the solute when the volume of the solute, $v$, is large.(b) Fits (solid lines) of Eq.~\ref{eq:Helfrich1} and Eq.~\ref{eq:Helfrich2} to estimates of solvation free energies per unit area of ideal hydrophobic spheres, and cylinders from the coarse grained lattice model. These fits are used to extract the macroscopic interfacial properties, $\kappa$, $c_0$ and $\delta$ (see text for description). $\Gamma$ is estimated by extrapolating estimates of $F/A$ for cubes from the lattice model to $\lim 1/L \to \infty$.}
\end{figure*}

The
basic physics emerging from these nonlinearities has been plumbed in
other contexts~\cite{Weeks1973,Chui1978,Kardar2007}. Here we exploit and extend the resulting understanding
and its connection with the molecular physics of microscopic density
fluctuations. We find that an appreciation of these issues constrains
but ultimately \textit{simplifies} the theoretical task of spanning
diverse length scales. The least complicated realization of the LCW
perspective,
involving no unknown parameters, can in fact suffice to describe
quantitatively the solvation of hydrophobic objects with various
shapes and sizes (see Fig.~\ref{fig2:results}). This
success establishes a minimally complicated model for the hydrophobic
effect that is faithful to the intrinsic softness of the air-water
interface, and should be useful to investigate solvation behaviors in various heterogenous environments~\cite{Setny2013,Patel2012}. It further allows us to parse contributions from various
length scales to material properties of the liquid-vapor interface, such
as
the
Tolman length, bending rigidity, and
spontaneous curvature.

The stability of a macroscopic liquid-vapor interface originates in
the statistical mechanics of phase transitions. Classic descriptions
of long-wavelength variations in that context include phenomenological
theories for smooth fields~\cite{Rowlison2002} and schematic lattice
models that implicitly coarse grain over scales smaller than a lattice
spacing $l$~\cite{Weeks1977,ReintenWolde2001}. The latter involve a minimum of parameters (as few
as $l$ and an energy scale $\epsilon$ of microscopic cohesion)
and are particularly convenient for numerical simulation. We, like others~\cite{Weeks1977,ReintenWolde2001},
therefore focus on such a discrete representation, 
where 
$n_i$
indicates the molecular density within lattice cell ${i}$ 
in a binary way: $n_i=1$ and $n_i=0$ denote locally
liquid-like and vapor-like density, respectively. Fluctuations of
these occupation variables are governed by a lattice gas Hamiltonian,

\begin{equation}
\label{eq:Ising}
H= -\epsilon
\sum_{\langle i,j\rangle} n_i n_j -\mu 
\sum_i
n_i\,, 
\end{equation}
where $\sum_{\langle i,j\rangle}$ denotes a sum over nearest neighbor
cells and $\mu$ is the chemical potential.  We have in mind systems
like ambient water that are close to coexistence, $\mu \approx
-3\epsilon +\Delta Pl^3$, where $\Delta P \gtrsim T/l^3$ is
the difference between ambient pressure and the liquid's vapor
pressure, and $T$ is given in units of $k_{\rm B}$.

How the parameters $l$ and $\epsilon$ should be assigned for a
particular material is a surprisingly subtle and pivotal
issue. Previous work has argued that $l$ should correspond to the
correlation length of density fluctuations in the liquid phase~\cite{ReintenWolde2001}. We are
concerned with fluids far below their critical points, setting this
length scale slightly in excess of a molecular diameter 
($l \approx4$\AA$\,$
in liquid water). 
The energy scale $\epsilon$ was inferred from
the cost of creating an interface 
at zero temperature. Equating this cost with the free energy per unit
area $\Gamma$ of a real liquid interface at finite temperature yields~\cite{ReintenWolde2001}
\begin{equation}
\label{eq:oldscaling}
\Gamma=\frac{\epsilon}{ 2 l^2}\,.
\end{equation} 
According to these arguments,
$\epsilon \approx 6.0   T$, 
in the case of water at ambient conditions.
We will show that 
such a high
value is problematic.

While this reasoning is sensible, it neglects entirely the influence
of interfacial fluctuations. To account for these shape variations, we
consider a macroscopically planar, 
fluctuating
interface between liquid and vapor
phases, within the so-called Solid-on-Solid (SOS) limit~\cite{Temperley:1952,Chui:1981,Fisher:1984,Nelson:2004,Weeks1977}.  In this
approximation the two phases are each assumed to be internally
homogeneous: $n_i=1$ everywhere in the liquid phase and $n_{i}=0$ everywhere in vapor, as roughly expected far from criticality.
Any configuration of this sort can be specified by the height $h_i$ of
the liquid phase in each column $i$ of the lattice (taking the
interface to be horizontal, with liquid below). At coexistence
the Hamiltonian can thus be rewritten as~\cite{Chui:1981}  
\begin{equation}
\label{eq:barehamil}
{H_0}= \frac{\epsilon}{4} \sum_{\langle i,j\rangle}|h_i-h_{j}|  \,. 
\end{equation}

Our strategy is to estimate the spectrum of capillary waves for the
SOS model (at finite temperature), and enforce agreement with that of
water.  According to capillary wave theory, the Fourier modes
$\hat{h}({\bf q})$ of a continuous interface fluctuate with squared
amplitude $\langle |\hat{h}({\bf q})|^2\rangle \propto 1/ \gamma_{\rm
  cap} q^2$, where $\gamma_{\rm cap}$ denotes the surface
tension governing capillary fluctuations. Molecular
simulations~\cite{Varilly2011} and experiments~\cite{Aarts2004}
indicate that modes of the air-water interface with wavelength $2\pi /
q \gtrsim 1$nm indeed follow this scaling, and that $\gamma_{\rm cap}$
corresponds closely with the thermodynamic surface tension
$\Gamma$~\cite{Mittal2008,Varilly2011}.  
Capillary wave scaling does not
necessarily hold for the lattice gas model or the SOS model, whose
interfacial roughness depends on temperature in
subtle ways~\cite{Weeks1973,Mon1989}.

In order 
to relate
$\gamma_{\rm cap}$ and $\epsilon$ for the SOS model, we ignore for the moment the discrete nature of fluctuations. 
We
examine interfacial statistics of the SOS model by
seeking the most representative Gaussian model $H _{\rm cap}= \frac{\gamma_{\rm cap}}{4} \sum_{\langle i,j\rangle} (h_i-h_{j})^2$. According to the Gibbs's variational principle,
\begin{equation}
\label{eq:variation}
F_0\leq F_{\rm cap}+\langle H_0-H_{\rm cap}\rangle_{\rm cap}\,,
\end{equation}
where $\langle\dots\rangle_{\rm cap}$ denotes an average taken with respect to the Hamiltonian $H_{\rm cap}$, $F_{\rm cap}$ denotes the free energy corresponding to $H_{\rm cap}$, and $F_0$ denotes the free energy corresponding to the SOS Hamiltonian $H_0$. 
This bound yields an optimal parameterization (see SM~\cite{Suppref1} for derivation)
\begin{equation}
\label{eq:centralresult1}
\beta \gamma_{\rm cap}=\frac{(\beta\epsilon)^2}{\pi}\,,
\end{equation}
where $\beta=1/T$. 
This result, scaling quadratically with $\epsilon$, is clearly
distinct from the low-temperature relationship in
Eq.~\ref{eq:oldscaling}.  We performed numerical simulations of the SOS model with values of $\epsilon/T$ in the range $0.4\leq\epsilon/T\leq 1.6$ and found  
the variational estimate to be very accurate for $\epsilon/T\lesssim1.4$ (see
Fig.~\ref{fig1:SOSlattice} and SM~\cite{Suppref1}). Comparison with 
the
lattice gas is favorable over a
more limited range, since the SOS approximation breaks down at low
values of $\epsilon$
as the critical point $\epsilon_c/T \approx
0.89$ is approached.  Specifically, the
range over which the capillary surface tension predicted by Eq.~\ref{eq:centralresult1} mirrors that
of the lattice gas is roughly bounded on the lower end by $\epsilon \gtrsim1.25  T $. We obtain this lower bound by computing $\gamma_{\rm cap}$ for the lattice gas model at coexistence~\cite{Hasenbusch1993} and comparing these estimates to those predicted by Eq.~\ref{eq:centralresult1}. The details of the calculation are presented in the SM~\cite{Suppref1}.
We will argue that the range, $1.25 \lesssim
\epsilon/T\lesssim 1.40$, is ideal for representing
liquid-vapor
interfaces. 

The breakdown of our variational estimate at high values of
$\epsilon/T$ reflects a well-known singularity in the statistical
physics of discretely fluctuating surfaces~\cite{Weeks1973,Chui1978}. Above a critical
value, $\epsilon_{\rm R}/T\approx1.63$, roughness of the SOS surface is
markedly suppressed~\cite{Weeks1973}. This transition to a quiescent interface, which
lacks capillary wave scaling, is generic to models that feature a
minimum energetic penalty for local deviations from flatness~\cite{Weeks1973,Chui1978}. 
Our variational estimate breaks down for values of $\epsilon/T$
noticeably below the roughening value. For values of $\epsilon/T$ below 
but close to
the roughening transition, the discrete constraints on lattice
fluctuations, which we have ignored while deriving
Eq.~\ref{eq:centralresult1}, become relevant and $\beta \gamma_{\rm
  cap}\neq (\beta\epsilon)^2/(\pi)$. Their effects can approximately
be assessed by adding a potential $V_0\equiv -2 y_0 \epsilon \sum_i
\cos(2\pi h_i)$ to the SOS Hamiltonian, Eq.~\ref{eq:barehamil}, with
continuous height fluctuations.  This potential penalizes
configurations in which height fluctuations deviate from discrete
values~\cite{Chui1978} with the constant $y_0$ determining the
strength of this penalty. Standard methods~\cite{Kardar2007,Chui1978}
can be used to calculate the corresponding renormalized surface
tension, $\gamma_{\rm cap}$. We choose $y_0$ so that the value of
$\epsilon/T$ at which this modified system undergoes a roughening
transition is close to that of the SOS lattice. We then find that for
$\epsilon/T \lesssim 1.4$, there is no significant renormalization due
to the discrete constraints. These calculations are described in the
SM~\cite{Suppref1}.

Based on this analysis we argue that the range of lattice gas
parameters consistent with the physics of hydrophobic solvation is
quite narrow. Large hydrophobic objects induce local drying,
generating microscopic analogs of a macroscopic interface between
liquid and vapor. Faithfully capturing fluctuations of such
microscopic interfaces requires that $\epsilon$ be smaller than the
critical value for roughening of the lattice gas interface, $\epsilon
<\epsilon_{\rm R}$~\cite{Weeks1973}.  
For values of $\epsilon$ slightly below $\epsilon_{\rm R}$,
the discrete nature of the lattice does not entirely suppress
long-wavelength capillary modes, but it nonetheless significantly influences
the statistics of surface fluctuations.
For example, the surface tension of the lattice gas is anisotropic in 
this regime~\cite{Mon1989}, depending
on
the orientation of the interface with respect to the axes of the cubic
lattice.  The implications of this 
and related lattice artifacts for 
solvation behaviors of
convex objects on cubic lattices have been discussed
previously~\cite{Varilly2011,Schrader2009a}. 
They encourage
using lattice coupling energies that are weaker still,
$\epsilon/T\lesssim1.4$, for which 
discreteness is an unimportant feature.

The SOS approximation, which
relies upon spatial uniformity  
within each phase, 
is well motivated for liquid water at ambient
conditions. Here, and in most liquids near their triple points,
spontaneous density fluctuations away from the average bulk value
$\rho_l$ are typically small even on molecular length scales. The range of cohesive energies $ 1.25   T \lesssim \epsilon \lesssim 1.4   T$ for which the variational estimate in Eq.~\ref{eq:centralresult1} is faithful is hence also optimal to  
represent fluctuations in real, far-from-critical
liquids. Through Eq.~\ref{eq:centralresult1}, this range of cohesive energies 
implies a correspondingly narrow range of 
appropriate lattice spacing $l$. In the case of water, using the experimental value of surface tension~\cite{Varilly2011}, $\gamma_{cap}\approx17.4   T/{\rm nm}^{2}=\epsilon^2/(\pi l^2)$, this
coarse graining length should, according to our arguments, lie
between 1.7 and 1.9~\AA. Some previous work has adopted values of
$l$ close to this range~\cite{Willard2009}, but in each case has assigned 
an energy scale through the low-temperature relationship Eq.~\ref{eq:oldscaling}.

Having tightly constrained the possible choices of $\epsilon$ and
$l$, we focus on implications for the theory and modeling of
aqueous solutions:
Can a suitably parameterized lattice model for long wavelength
variations in density, together with a simple theory for molecular
scale fluctuations, accurately predict nontrivial solvation behavior?
To do so, we employ the LCW perspective in its simplest incarnation,
put forth in Ref~\cite{Lum1999}. Locally liquid-like regions ($n_i=1$) additionally support short-wavelength density fluctuations which are assumed to obey Gaussian statistics~\cite{Chandler1993}, as
characterized by the two-point correlation function $\chi({\bf r}-{\bf
  r}^\prime)=\rho_l l({\bf r}-{\bf r}^\prime)+\rho_l^2(g({\bf
  r}-{\bf r}^\prime)-1)$, where $g(r)$ denotes the radial distribution function~\cite{Rowlison2002}.

We consider solutes that are ideally
hydrophobic, whose sole influence is to exclude solvent from a volume
$v$. The constraint of solvent evacuation within a lattice cell $i$
can be accommodated either through variation in the 
slowly varying density
field ($n_i=0$), or else through a variation in the above mentioned short wavelength Gaussian density field in a locally
liquid-like region. Integrating out short-wavelength fluctuations
yields an effective Hamiltonian for lattice occupation variables in
the presence of a solute~\cite{ReintenWolde2001,Varilly2011}:
\begin{eqnarray}
\label{eq:LCW}
&&H_v[n_i]= -\epsilon\sum_{\langle i,j\rangle}
n_i n_j -\mu \sum n_i +  T \left[\frac{
    N_v^2}{2\sigma_v}\right] +\frac{T C}{2}\,, \\
\label{eq:LCWdetails1}
&& N_v=\Sigma_i \rho_l n_i v_i\, , \sigma_v=\int_{{\bf r}\in v}\int_{ {\bf r}^\prime\in v}  \Theta({\bf r}) \chi({\bf r},{\bf r}^\prime) \Theta({\bf r}^\prime)\,,
\label{eq:LCWdetails2}
\end{eqnarray}
where 
\begin{equation}
  C=\begin{cases}
    \ln\left(2\pi \sigma_v\right) & \text{if $\langle N\rangle_v>1$},\\
    \rm{max}\left[\ln\left(2\pi\sigma_v\right),\langle N\rangle_v\right] & \text{otherwise}\,,
  \end{cases}
\end{equation}
and $\Theta ({\bf r})=1$ if the lattice cell containing ${\bf r}$ is
occupied and vanishes otherwise, and $v_i$ is the volume of overlap
between $v$ and lattice cell $i$. The coarse-grained model defined by
Eq.~\ref{eq:LCW} includes as free parameters only the energy  
and
length 
scales of the underlying lattice gas, which we set
as $\epsilon = 1.35   T $, and consistent with the statistical mechanics of rough interfaces (Eq.~\ref{eq:centralresult1}), $l = 1.84$~\AA~(see inset, Fig.~\ref{fig1:SOSlattice}). We used $\mu=-3\epsilon+1.51 \times 10^{-4} T$~\cite{Willard2009} for these simulations. Note that the only inputs to this theory are the surface tension of water, and its pair correlation $g(r)$~\cite{Rowlison2002}. 

Using Monte Carlo simulations of this coarse-grained description, we
computed the reversible work $F$ required to transfer volume-excluding
solutes from vapor into the bulk liquid phase (details of implementation in SM~\cite{Suppref1}).  We
focus first on spherical hydrophobes, for which solubility has been
previously determined as a function of radius $R$ from detailed
molecular simulations~\cite{Huang2001}. The free energy per unit solute surface area $A
= 4\pi R^2$ is plotted in Fig.~\ref{fig2:results}(1). Results for the lattice model agree
very well with simulation data. As a more stringent test, we computed solvation free energies for
hydrophobic objects that vary not only in scale but also in
shape. It is not obvious whether a microscopic sphere with given
surface area should differ substantially in solubility from, e.g., a
cube with the same area. As shown in Fig.~\ref{fig2:results}, solvation properties can
in fact be 
very 
sensitive to such geometric details. For all of the shapes
we considered (spheres, cubes, and cuboids), $F/A$ grows rapidly as
radius $R$ (or edge length $L$ for the cubes and cuboids considered here) grows to $\sim 1$nm, then increases much more gradually at larger
$R$ (or $L$). The details of this dependence, however, differ significantly. A
naive extrapolation could even suggest that $F/A$ approaches different
limiting values at large $R$ for different shapes. These behaviors are
observed in molecular simulations (using the SPC/E model of water,
details in SM~\cite{Suppref1}) and for the coarse-grained theory of Eq.~\ref{eq:LCW}, with
remarkably close correspondence between the two approaches. From the
solubility of cubic volumes, we extract the thermodynamic surface
tension $\Gamma =14.92\,   T/{\rm nm}^{2}$  of our lattice model by
extrapolating to $L^{-1}\rightarrow 0$ (see Fig.~\ref{fig3:results}). 

The sensitivity of a hydrophobe's solubility to its shape reflects
properties of liquid-vapor interfaces that are more subtle than
macroscopic surface tension (i.e., $\Gamma$ or $\gamma_{\rm cap}$).
Corresponding material parameters are conventionally defined in terms
of an expansion in powers of curvature $1/R$. The Tolman length
$\delta$, 
for example, is defined according to $F_{sph}/A =
\Gamma(1-2\delta/ R+\dots)$.
A more directly mechanical view is
provided by mapping this 
curvature-dependent response
onto that of an elastic
sheet. According to Helfrich's phenomenological theory, the elastic
free energy of a thin shell with bending rigidity $\kappa$ and
spontaneous curvature $c_0$ is given by~\cite{Sedlmeier2012}
\begin{eqnarray}
\label{eq:Helfrich1}
&&\frac{F_{sph}}{A}=\Gamma-\frac{4 \kappa c_0}{R}+C_1\frac{1}{R^2}+C_2 \frac{1}{R^3}\,, \\ 
\label{eq:Helfrich2}
&& \frac{F_{cyl}}{A}=\Gamma -\frac{2 \kappa c_0}{R}+\frac{\kappa}{2 R^2}+D_1 \frac{1}{R^3}\,, 
\end{eqnarray}
for spherical and cylindrical shapes, respectively. ($C_1$, $C_2$ and
$D_1$ are constants determining still more subtle material properties
that are not discussed here.) Simultaneously fitting theoretical results for
spherical and cylindrical solutes to the form of Eqs.~\ref{eq:Helfrich1},\ref{eq:Helfrich2} (see Fig.~\ref{fig3:results})
yields $\kappa=-3.54  T$ and $c_0=0.29 nm^{-1}$, and therefore
$\delta=2\kappa c_0/\Gamma=-0.14 nm$. These values are consistent
with previous estimates from molecular simulation~\cite{Sedlmeier2012}.

The ability of such a coarse-grained theory to capture the precise
shape dependence of interfacial thermodynamics is striking, given the
minimal molecular detail Eq.~\ref{eq:LCW} add to the generic lattice
gas. The lattice model, by itself, cannot in fact produce interfaces with
nonzero spontaneous curvature $c_0$, a consequence of symmetry between
liquid and vapor phases in this crude description~\cite{Fisher1984}.  Accurately
predicted nonzero values of $c_0$ and $\delta$ therefore emerge entirely
from the coupling of Gaussian density fluctuations at small scales to
the generic interfacial roughness of an Ising model. These mechanical
details of the air-water interface, which might appear to reflect
geometric intricacies of hydrogen bonding, are thus encoded in the
simplest measure of microscopic structure in the bulk liquid, its pair correlation function $g(r)$.

We gratefully acknowledge extremely useful discussions with John D Weeks,  Gerhard Hummer, David Chandler, and David Limmer. 
This project was supported by the US Department of Energy, Office of
Basic Energy Sciences, through the Chemical Sciences Division (CSD) of the Lawrence Berkeley National Laboratory (LBNL), under Contract DE-AC02-05CH11231.

%

\end{document}


\title{Supplemental Material for\\
``Putting water on a lattice: The importance of long wavelength density fluctuations in theories of hydrophobic and interfacial phenomena ''}
\author{Suriyanarayanan Vaikuntanathan}
\affiliation{Material Sciences Division, Lawrence Berkeley National Lab, 
Berkeley, CA 94720}
\author{Phillip L. Geissler}
\affiliation{Material Sciences Division, Lawrence Berkeley National Lab, 
Berkeley, CA 94720}
\affiliation{Department of Chemistry, University of California, Berkeley, 
CA 94720}
\affiliation{Chemical Sciences Division, Lawrence Berkeley National Lab, 
Berkeley, CA 94720}

\maketitle

\section{Fluctuations in the SOS model}
The Solid on Solid (SOS) system is described by the Hamiltonian
\begin{equation}
\label{SIeq:SOS}
H_0=\frac{\epsilon}{4}\sum_{\langle i,j\rangle } |h_i-h_{j}|\,.
\end{equation} 
As described in the main text, we ignore the discrete nature of fluctuations in SOS Hamiltonian above and
seek the most representative Gaussian model with Hamiltonian, $H _{\rm cap}= \frac{\gamma_{\rm cap}}{4} \sum_{\langle i,j\rangle} (h_i-h_{j})^2$. According to the Gibbs's variational principle,
\begin{equation}
\label{SIeq:variation}
F_0\leq F_{\rm cap}+\langle H_0-H_{\rm cap}\rangle_{\rm cap}\,,
\end{equation}
where $\langle\dots\rangle_{\rm cap}$ denotes an average taken with respect to the Hamiltonian $H_{\rm cap}$, $F_{\rm cap}$ denotes the free energy corresponding to $H_{\rm cap}$, and $F_0$ denotes the free energy corresponding to the SOS Hamiltonian $H_0$ (with continuous fluctuations). We find $\gamma_{\rm cap}$ by setting 
\begin{equation}
\label{SIeq:varsetup}
\frac{{\partial F_{\rm cap}}}{{\partial \gamma_{\rm cap}}}+\frac{{\partial }}{{\partial \gamma_{\rm cap}}}\langle H_0\rangle_{\rm cap}=0
\end{equation}
To proceed, we consider $P_{\rm cap}(z)$, the probability density associated with observing a height difference $z$ between a particular pair of nearest neighbors, $i$ and $j$, in the Gaussian ensemble, 
\begin{equation}
\label{SIeq:inter0}
P_{\rm cap}(z)=\langle \delta(h_i-h_j-z)\rangle_{\rm cap}=\frac{1}{2\pi}\int d k \langle e^{-\rm{i} k (h_i-h_j)}\rangle_{\rm cap}e^{\rm{i} k z}  \,.
\end{equation}
To evaluate this density, we set the origin of the two dimensional plane perpendicular to the interface, denoted by $\hat 0$, at the site $j$ and without loss of generality assume that site $i$ is one lattice unit away on the positive x-axis. Denoting the unit vector along the x-axis by $\hat{x}$, we obtain the following identity (after some tedious but straightforward algebra),  
\begin{equation}
\label{SIeq:inter1}
\langle e^{-\rm{i} k (h_{\hat x}-{h_{\hat 0}})}\rangle_{\rm cap}=\exp\left[\frac{-k^2}{2 L^2}\sum_{{\bf q}}\frac{2-2\cos\left({\bf q}\cdot {\hat x}\right)}{\gamma_{\rm cap}\left(4-2 \cos({\bf q}\cdot \hat x)-2\cos({\bf q}\cdot \hat y)\right)}\right]\,,
\end{equation}
where ${\bf q}$ denotes a (two dimensional) wave vector, $L^2$ is the number of lattice sites, and $\sum_{\bf q}$ denotes a sum over the wave vectors. Using the symmetry of the lattice in ${\bf x}$ and ${\bf y}$ directions, the summation in Eq.~\ref{SIeq:inter1} works out to 
\begin{equation}
\label{SIeq:inter1}
\sum_{{\bf q}}\frac{2-2\cos\left({\bf q}\cdot {\hat x}\right)}{\left(4-2 \cos({\bf q}\cdot \hat x)-2\cos({\bf q}\cdot \hat y)\right)}=\sum_{{\bf q}}\frac{2-2\cos\left({\bf q}\cdot {\hat y}\right)}{\left(4-2 \cos({\bf q}\cdot \hat x)-2\cos({\bf q}\cdot \hat y)\right)}=\frac{L^2}{2}\,,
\end{equation}
where we have used the fact that the summation has $L^2$ terms. 
We can now compute the density using Eq.~\ref{SIeq:inter0} as 
\begin{equation}
P_{\rm cap}(z)=\sqrt{\beta \gamma_{\rm cap}}\frac{e^{-\beta \gamma_{\rm cap} z^2}}{\sqrt \pi}\,.
\end{equation}

This allows us to compute $\langle |h_i-h_j|\rangle$,  
\begin{equation}
\langle |h_i-h_j|\rangle=2 \int_0^\infty z P_{\rm cap}(z) dz= \frac{1}{\sqrt{\pi\beta \gamma_{\rm cap}}}\,.
\end{equation}

The variational calculation described in Eq.~\ref{SIeq:varsetup} now gives us  
\begin{equation}
\frac{\beta\epsilon}{\sqrt{\pi}\left(\beta\gamma_{\rm cap}\right)^{3/2}}=\frac{1}{\beta\gamma_{\rm cap}}\,,
\end{equation}
which yields 
\begin{equation}
\label{SIeq:central}
\beta \gamma_{\rm cap}=\frac{(\beta\epsilon)^2}{\pi}\,.
\end{equation}

\section{Importance of discrete constraints}
The effects of discrete constraints on Eq.~\ref{SIeq:central}
can approximately be assessed by constructing a
Sine-Gordon like Hamiltonian~\cite{Chui1978},
 \begin{equation}
\label{eq:barehamilSG}
{H_{\rm SG}}=  \frac{\epsilon}{4} \sum_{\langle i,j\rangle}|h_i-h_{j}|-2 y_0 \epsilon \sum_i \cos(2\pi h_i) \,,
\end{equation}
whose second summation penalizes configurations in which height
fluctuations deviate from discrete values. The constant $y_0$
determines the strength of 
this
penalty. The renormalized surface tension $\tilde \gamma_{\rm cap}={1}/{q^2 \langle |\hat
  h(q)|^2\rangle_{\rm SG}}$ can be computed using a perturbation theory in $y_0$, 
\begin{equation}
\label{SIeq:RG1}
\frac{1}{\langle |\hat h(q)|^2\rangle_{\rm SG}}=\frac{1}{\langle |\hat h(q)|^2\rangle_{\rm cap}}-\frac{O(y_0^2)}{\langle |\hat h(q)|^2\rangle_{\rm cap}}\,
\end{equation}
 where we have mapped the SOS Hamiltonian with continuous fluctuations onto the Gaussian reference Hamiltonian using Eq.~\ref{SIeq:central}, observed that the term linear in $y_0$ vanishes, and where  
\begin{equation}
O(y_0^2)=\frac{(2\epsilon_0 y_0)^2}{2}\left[\langle \sum_{i,j}\cos(2\pi h_i) \cos(2\pi h_j)|\hat h(q)|^2\rangle_{\rm cap}-\langle \sum_{i,j}\cos(2\pi h_i) \cos(2\pi h_j)\rangle_{\rm cap}\langle|\hat h(q)|^2\rangle_{\rm cap}\right]\,.
\end{equation}
Here, $\sum_{i,j}$ denotes a double summation over all lattice sites. After some tedious (but straightforward) algebra~\cite{Kardar2007,Chui1978}, Eq.~\ref{SIeq:RG1} can be rewritten as 
\begin{equation}
\label{eq:KTrecursion}
\frac{ \tilde \gamma_{\rm cap}}{T}=\frac{\gamma_{\rm cap}}{T}+ {{(2\beta\epsilon y_0\pi^{-\pi/\beta\gamma_{\rm cap}})}^2}\pi^3 \int_{1}^{\infty} dr \left(r\right)^{3-2\pi/ \beta\gamma_{\rm cap}}\, ,
\end{equation}
where we have assumed that number of lattice sites, $L^2$, is large, replaced the summation $\sum_{i,j}$ by integrals, and have used the approximation 
\begin{equation}
\label{SIeq:Besselapprox}
\int_0^{\pi r} \frac{1-{ {\rm J}_{0}(q)}}{q}\approx \ln \pi r \,,
\end{equation}
when $r>>1$. In Eq.~\ref{SIeq:Besselapprox}, ${ {\rm J}_{0}(z) }$ denotes the zeroth order Bessel function of the first kind. 

\begin{figure}[tbp]
\includegraphics[width=0.7\linewidth]{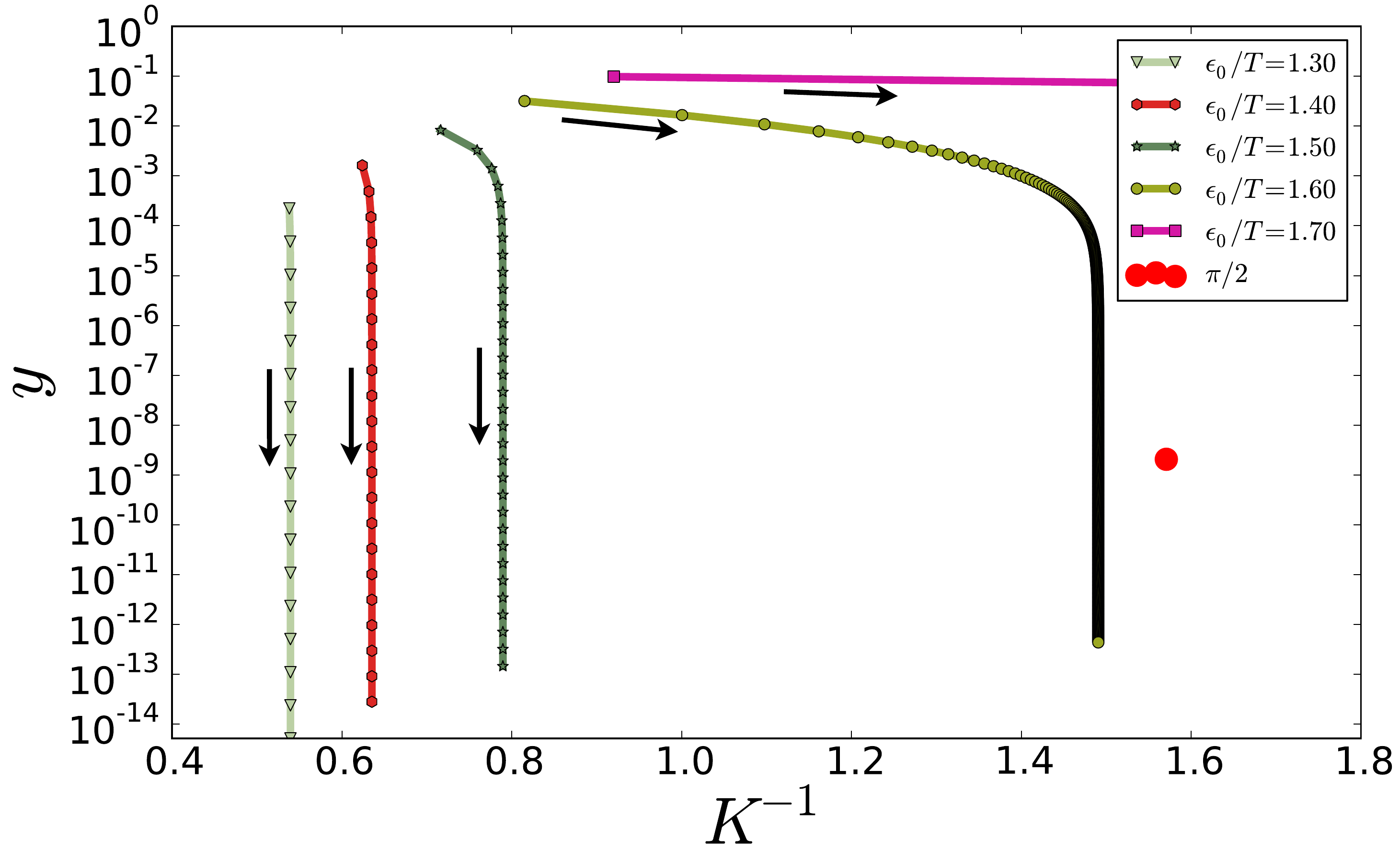}
\caption{Evolution of $K^{-1}\equiv \beta\gamma_{\rm cap}$ and $y$\textemdash this reflects the importance of the discrete constraints\textemdash under the renormalization flow equations. The arrows indicate the direction of the flow. For $\epsilon/T<1.4$, $y$ is driven to zero while the surface tension hardly gets renormalized.}
\label{SIfig:RGflow}
\end{figure}

This equation can be solved using the Kosterlitz-Thouless renormalization approach~\cite{Chui1978,Kardar2007}. The solutions of the renormalization group flow equations~\cite{Chui1978,Kardar2007} are described in Fig.~\ref{SIfig:RGflow} with the definitions $K^{-1}\equiv \gamma_{\rm cap}/T$, and $y\equiv {2\epsilon y_0}\pi^{(-\pi K)}/{T}$. The initial conditions for the flow equations were $K^{-1}= (\beta\epsilon)^2/\pi$, and $y_0=4.6$. For values $\epsilon/T<1.4$, there is no significant renormalization due to the discrete constraints and $\beta\gamma_{cap}\approx(\beta\epsilon)^2/\pi$. For the value of $y_0$ chosen here, the roughening transition is approximately at $\epsilon/T\approx 1.64$. Near this value of $\epsilon/T$, the asymptotic value of the surface tension is equal to theoretically predicted universal value, $\beta\gamma_{\rm cap}=\pi/2$~\cite{Chui1978,Kardar2007}.

\section{Verification of scaling predicted in Eq.~\ref{SIeq:central}}

\begin{figure}[tbp]
\includegraphics[width=0.7\linewidth]{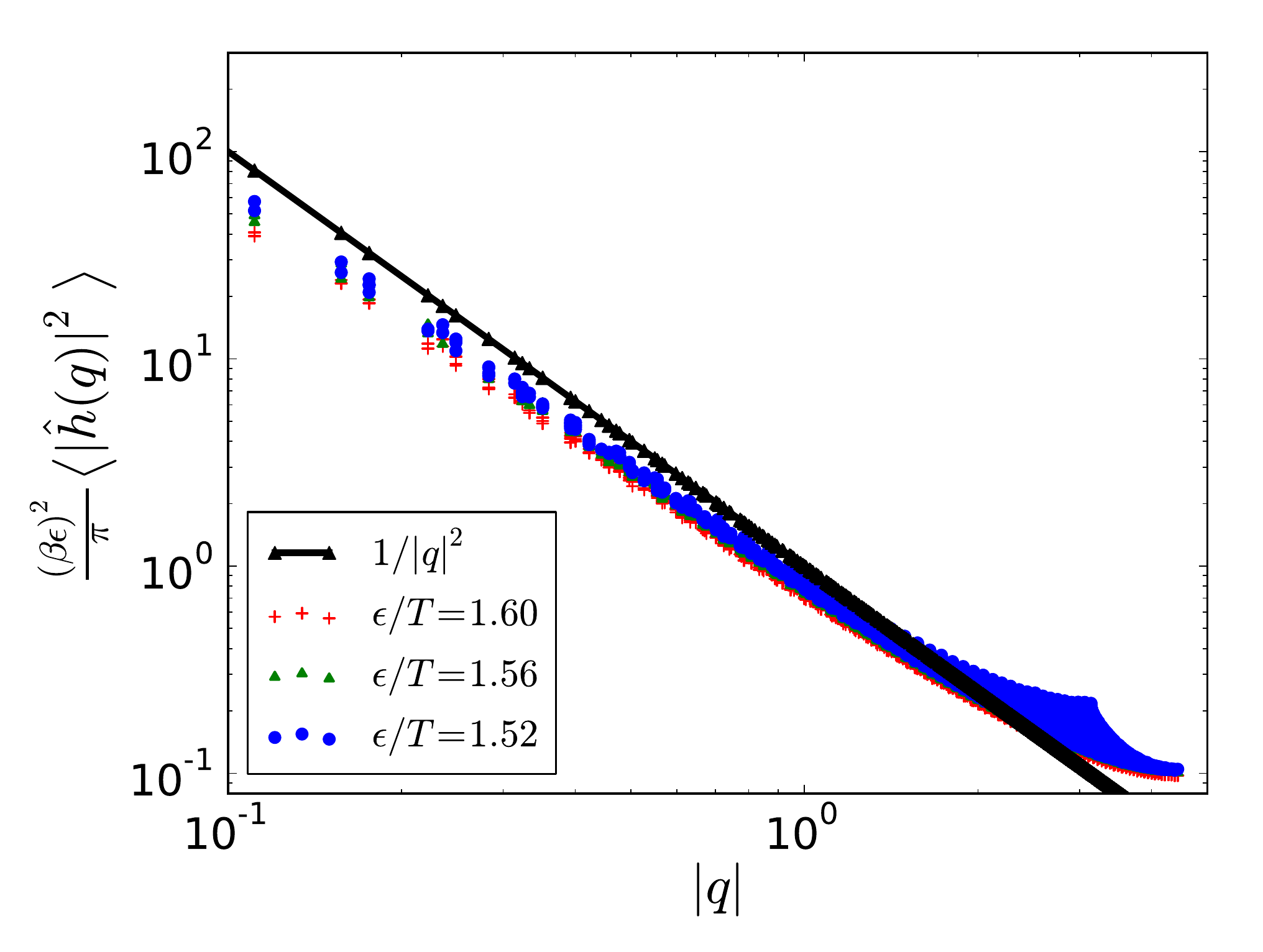}
\caption{Breakdown of variational approximation for $\epsilon\leq\epsilon_R$. Results for $(\beta\epsilon)^{2}\langle |\hat h (q)|^2\rangle/\pi$ from SOS simulations are plotted alongside the prediction $1/|q|^2$ from Eq.~\ref{SIeq:central}. For the values of $\epsilon/T$ considered here, the discrete nature of the lattice can no longer be ignored.}
\label{SIfig:SOS}
\end{figure}

As we demonstrated in the main text, the surface tension, $\gamma_{\rm cap}$, of the SOS lattice is well approximated by the relation in Eq.~\ref{SIeq:central} for values of $\epsilon/T$ approximately below $\epsilon/T\lesssim1.4$. In Fig.~\ref{SIfig:SOS}, we consider cases $1.5< \epsilon/T\leq 1.6$, which highlights the quantitative importance of lattice discreteness for values of $\epsilon$ even modestly below $\epsilon_R$. In particular, even though the SOS lattice supports capillary waves for these values of $\epsilon/T$, its surface tension is markedly different from that predicted by Eq.~\ref{SIeq:central}. 

\section{Estimates of surface tension of the liquid vapor interface in the Lattice model}
\begin{figure}[htbp]
              \subfigure[$\epsilon/T=1.20$]{
                \includegraphics[scale=0.35]{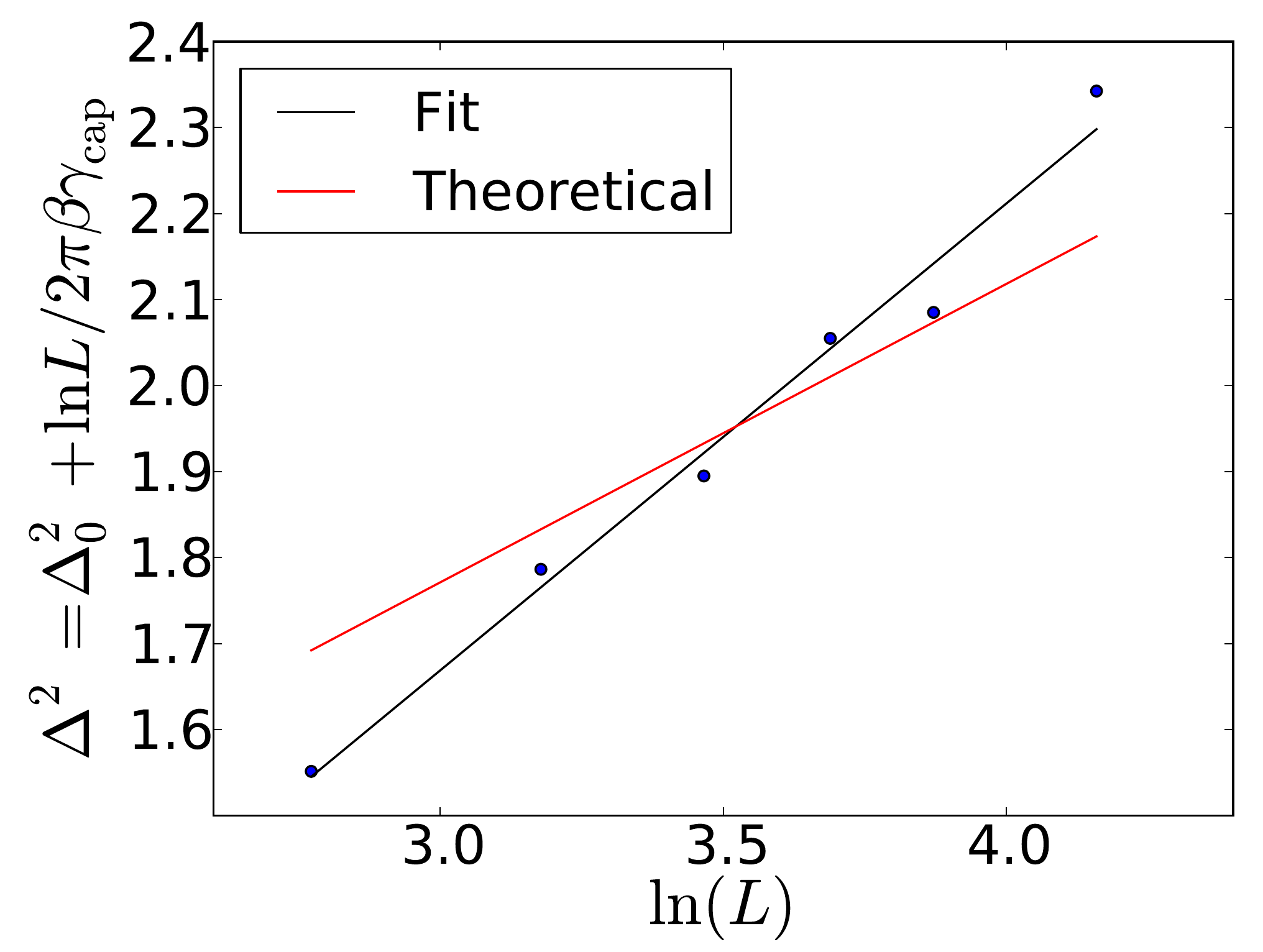}
                \label{fig1a:SOS}}
             \subfigure[$\epsilon/T=1.25$]{
                \includegraphics[scale=0.35]{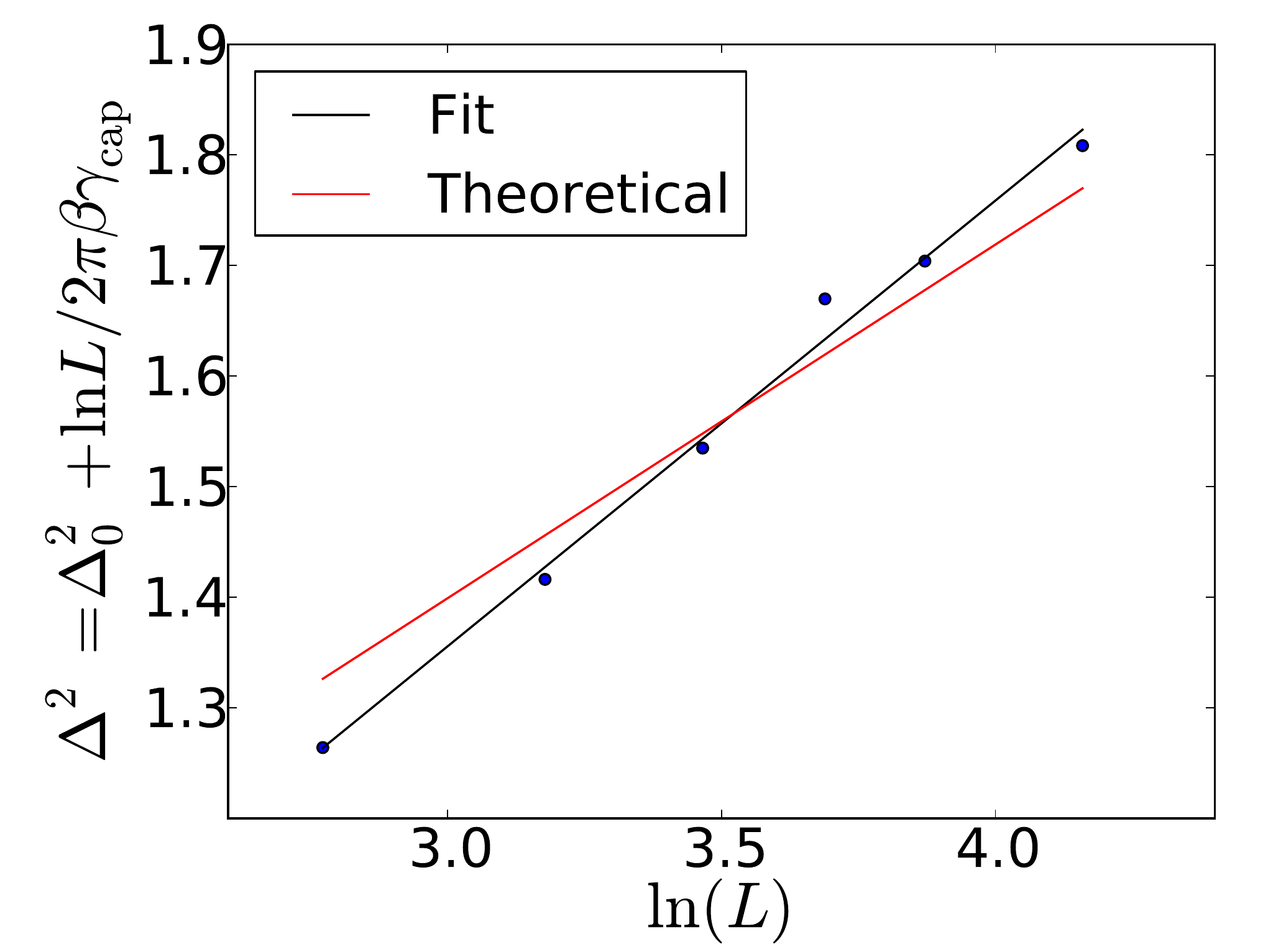}
                \label{fig1b:SOSlattice}}
       \subfigure[$\epsilon/T=1.30$]{
                \includegraphics[scale=0.35]{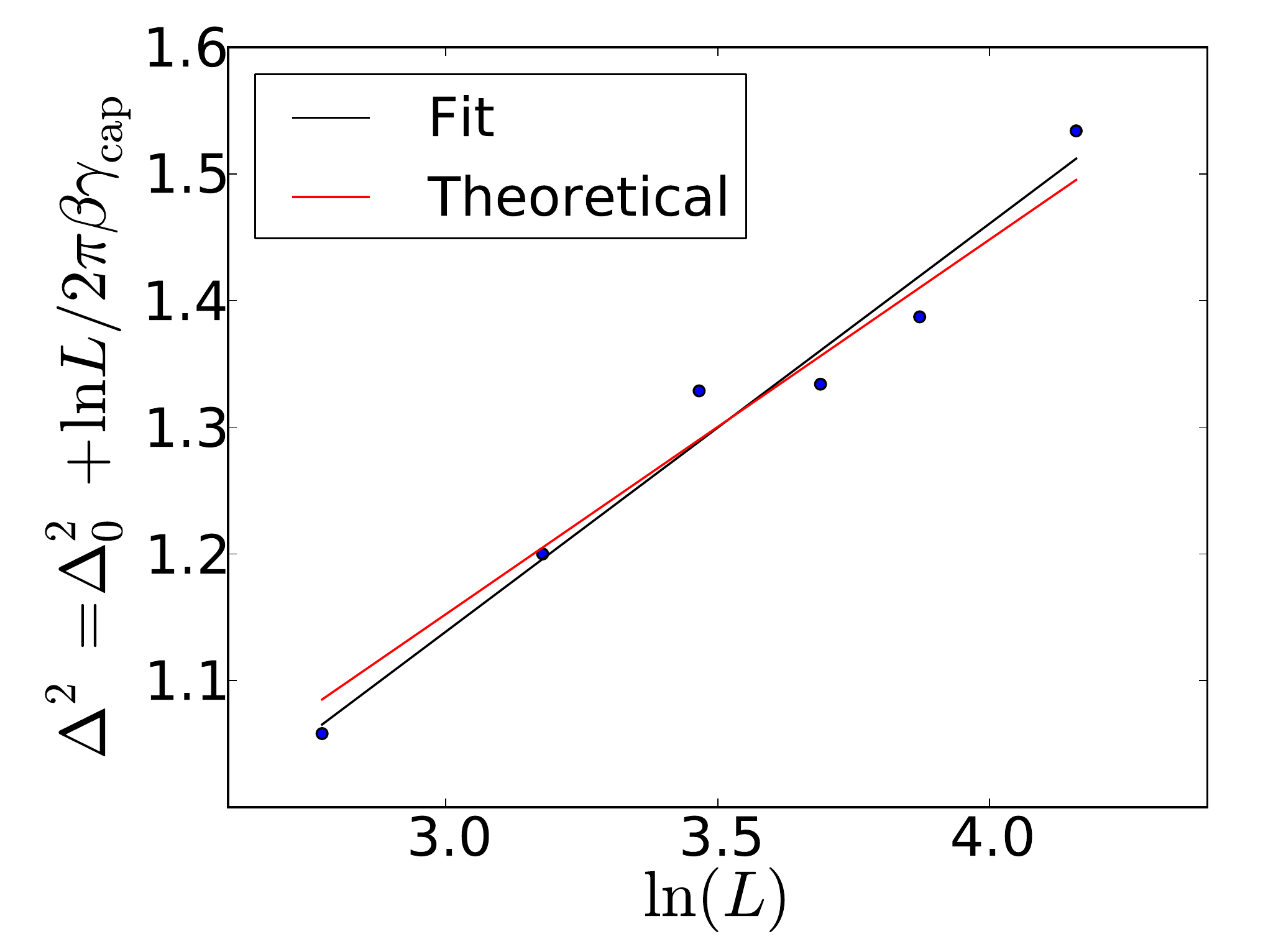}
                \label{fig1c:SOS}}
       \subfigure[$\epsilon/T=1.40$]{
                \includegraphics[scale=0.35]{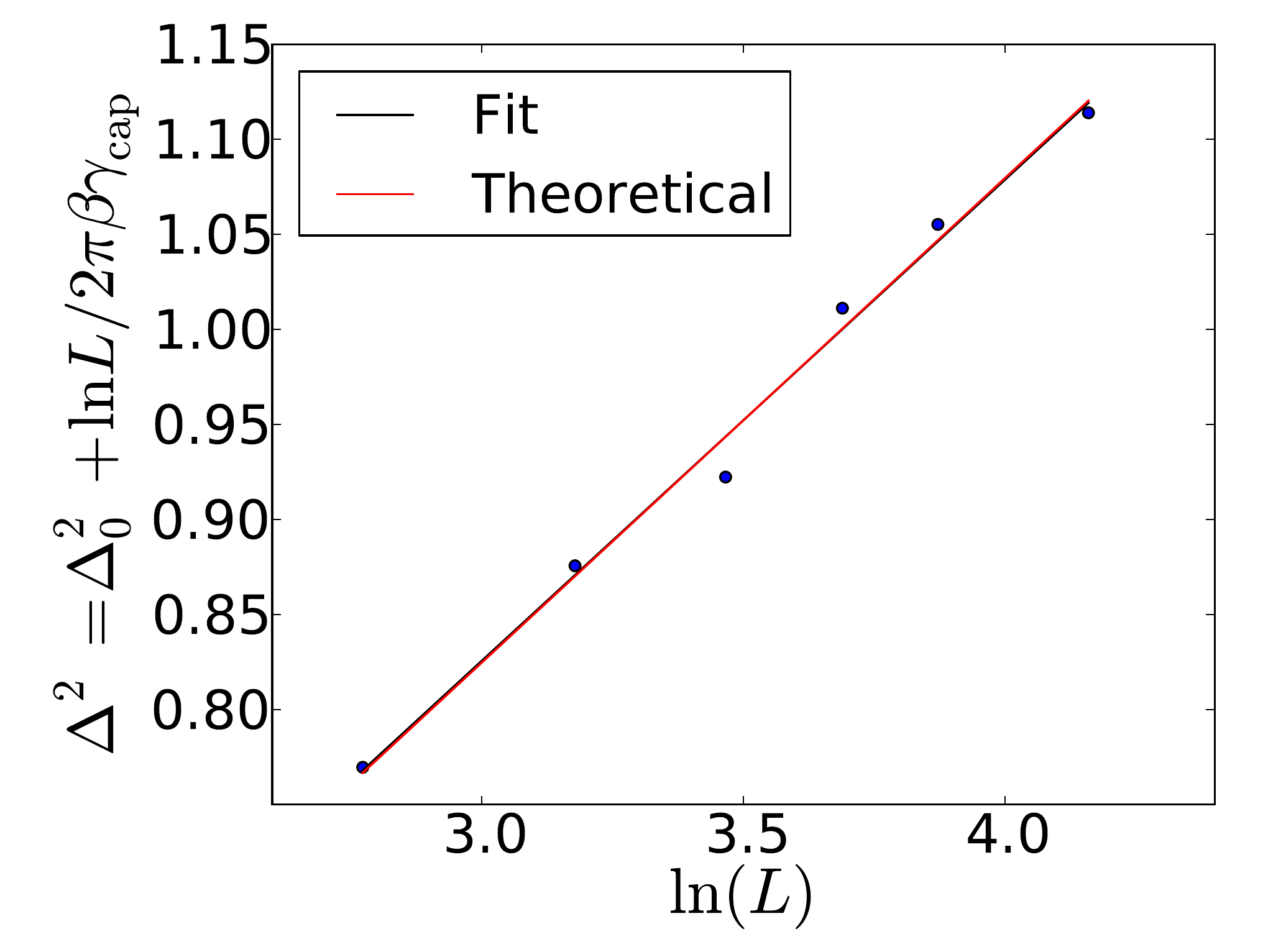}
                \label{fig1d:SOS}}   
       \caption{Scaling of the interface width as a function of the lateral dimension $L$ for a lattice gas at coexistence at various values of $\epsilon/T$. Eq.~\ref{SIeq:central} breaks down visibly as $\epsilon/T$ is lowered below $\epsilon/T=1.25$.}
       \label{SIfig:scaling}
\end{figure}

We computed the capillary surface tension of the lattice gas model at coexistence to verify the scaling predicted in Eq.~\ref{SIeq:central}. The interface width was estimated using the technique described in Ref.~\cite{Hasenbusch1993} and requires the use of a clustering algorithm that removes all bubbles from the bulk liquid and all droplets from the vapor phases. The surface tension $\gamma_{\rm cap}$ is obtained by studying the scaling of the interface width as a function of the lateral dimension, $L$, and using the capillary relation
\begin{equation}
\Delta^2=\Delta^2_0+\frac{1}{2\pi \beta \gamma_{\rm cap}} \ln{L}\,,
\end{equation}
where $\Delta$ denotes the interface width, and $\Delta_0$ is the intrinsic interface width. 
As we demonstrate in Fig.~\ref{SIfig:scaling} , Eq.~\ref{SIeq:central} provides an accurate estimate of $\gamma_{\rm cap}$ only in the range $1.25 \lesssim \epsilon/T \lesssim 1.40$.

\section{Implementation of the coarse grained model}
Implementation of the coarse grained model requires the computation of the integral 
\begin{equation}
\label{SIeq:g_int}
\sigma_v=\int_{{\bf r}\in v}\int_{ {\bf r}^\prime\in v}  \Theta({\bf r}) \chi({\bf r},{\bf r}^\prime) \Theta({\bf r}^\prime)\,,
\end{equation}
where $\chi({\bf r}-{\bf
  r}^\prime)=\rho_l \delta({\bf r}-{\bf r}^\prime)+\rho_l^2(g({\bf
  r}-{\bf r}^\prime)-1)$, and the other definitions are given in the main text. 
We approximate Eq.~\ref{SIeq:g_int} as 
  \begin{equation}
\label{SIeq:g_int}
\sigma_v\approx \sum_{i}\sum_j n_i b_v(i) \left[\int_{{\bf r}\in i}\int_{ {\bf r}^\prime\in j} \chi({\bf r},{\bf r}^\prime)\right]n_j b_v(j) \,,
\end{equation}
where $b_v(i)$ is the fraction of cell $i$ occupied by the solute. The integral in Eq.~\ref{SIeq:g_int} is performed using the tables provided in Ref.~\cite{Narten1971}.

\section{Details of Free energy calculations}
For free energy calculations on the coarse grained model, we solvate volume excluding solutes of various shapes and sizes, interpolating between a 
point solute and a final size of interest, and sequentially estimate
the free energy difference between neighboring solutes in this list
using the free energy perturbation identity~\cite{Frenkel2002}.  We use the Hamiltonian energy function described in the main text for these free energy calculations. 

In the case of spheres, we compare free energy estimates from the coarse grained model to estimates of solvation free energy of hard spheres in
SPC/E water~\cite{Berendsen1987} in Refs.~\cite{Huang2001,Chandler2005}. 
For the other shapes, we simulated 6912 SPC/E water molecules in a slab geometry in a NVT ensemble. The dimensions of the slab are 7.2 nm $\times$ 7.2 nm $\times$ 3.6 nm with liquid vapor interfaces above and below the slab. The simulation was performed at $T=300 K$. Particle Mesh Ewald was used to handle the electrostatic interactions and the SHAKE algorithm was used to constrain the bonds in the water. The simulations were performed using the LAMMPS package. We use the INDUS~\cite{Patel2010} umbrella sampling method to estimate the solvation free energies. The slab configuration was chosen for reasons described in Ref.~\cite{Patel2010}. 

In Fig.~\ref{SIfig:PV}, we plot estimates of the distribution $P_v(N)$, the probability distribution associated with finding $n$ water molecules in a probe volume $v$, obtained from our simulations for $v\sim$ cubes and $v\sim$ cuboids. The solvation free energy is given by $F=-\beta^{-1}\ln P_v(0)$.     

\begin{figure}
   \label{SIfig:PV}
                 \subfigure[]{
                \includegraphics[scale=0.35]{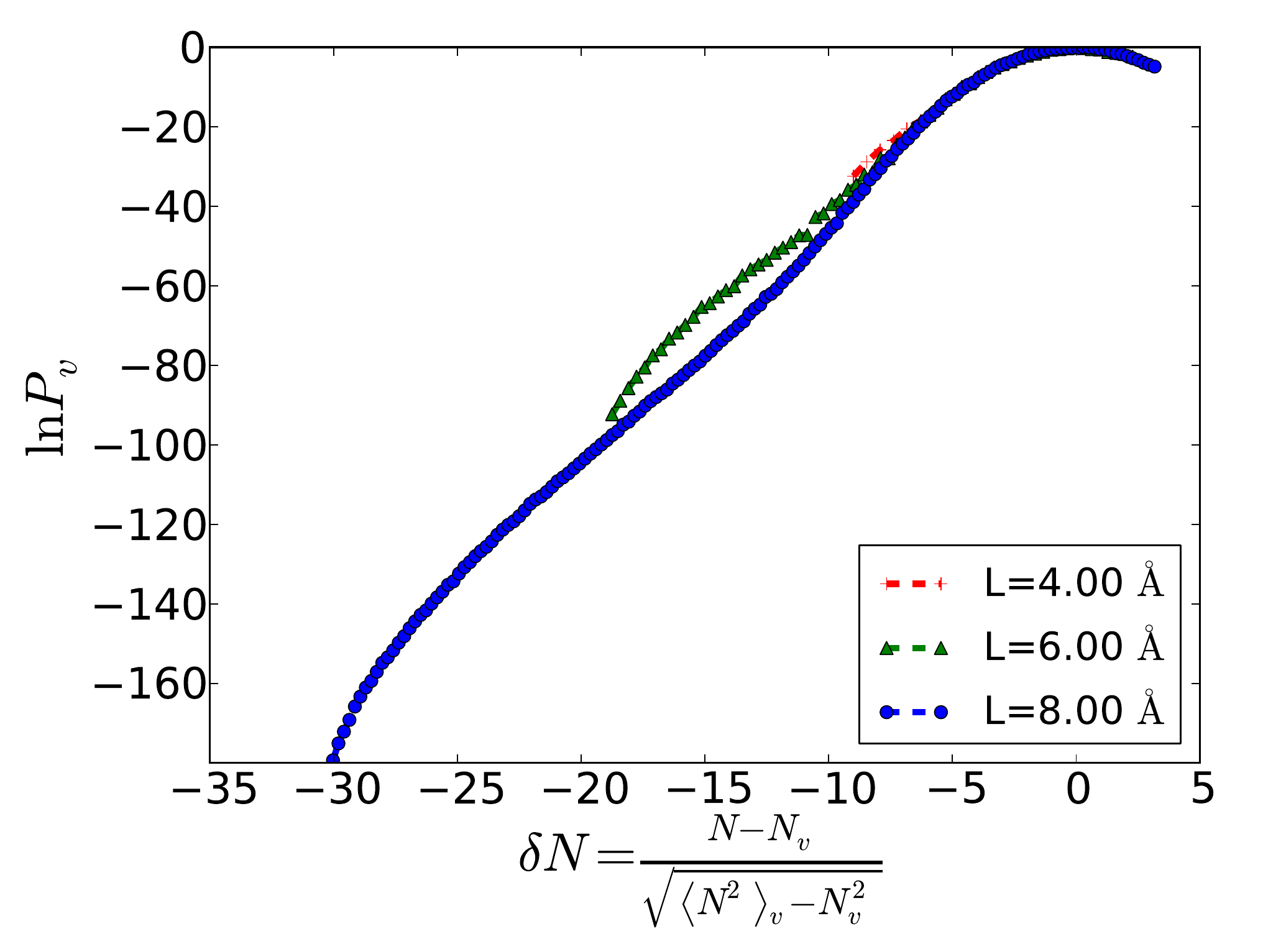}
                \label{fig2a}
     }
            \subfigure[]{
                \includegraphics[scale=0.35]{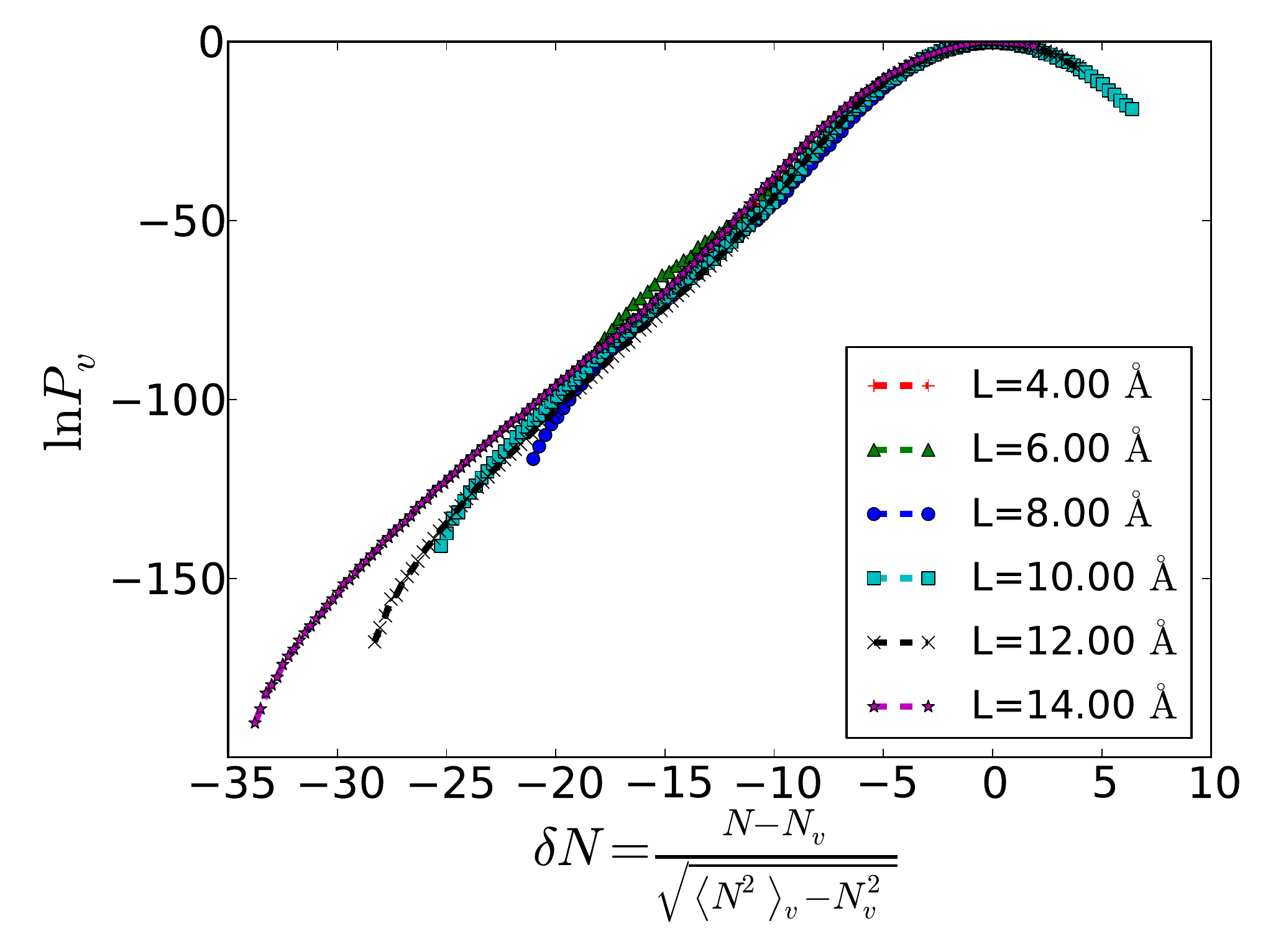}
                \label{fig2b}
 }
       \caption{Plots of $\ln P_v$ as function of $\delta N\equiv \frac{N-N_v}{\langle N^2\rangle_v-N_v^2}$, where $N_v$ denotes the average number of water molecules in the probe volume $v$. The probe volume $v$ is cuboid shaped ($L\times 12 \AA \times 12 \AA$) for the instances in Fig.~\ref{fig2a} and cube shaped for those in Fig.~\ref{fig2b}. See text for details of simulation. }
       \label{SIfig:PV}
\end{figure}


%